\documentclass[10pt,aps,prb,twocolumn,amsmath,amssymb,superscriptaddress]{revtex4-2}
\usepackage{amsmath}
\usepackage{graphicx,subfigure}
\usepackage{color}
\usepackage{multirow}
\usepackage[caption = false]{subfig}
\usepackage[normalem]{ulem}
\usepackage{epstopdf}
\usepackage{array}
\usepackage{algcompatible}
\usepackage[floatrow]{trivfloat}
\trivfloat{algorithm}

\usepackage{float}
\usepackage{placeins} 
\floatstyle{plaintop}
\restylefloat{table}

\def\beq{\begin{eqnarray}}
\def\eeq{\end{eqnarray}}
\def\beqq{\begin{eqnarray*} \color{blue} }
\def\eeqq{\end{eqnarray*}}

\def\V{\mathcal{V}}
\def\P{\mathcal{P}}

\newcommand{\bra}[1]{\left\langle #1 \right|}
\newcommand{\ket}[1]{\left|  #1  \right\rangle}

\newcommand{\brakett}[3]{\bra{#1} #2 \ket{#3}}

\def\basis{\mathcal{B}}
\def\psibasis{\Psi^{\basis}}

\def\kinop{\hat{T}}
\def\weeop{\hat{W}_{\text{ee}}}

\begin{document}

\title{Accurate energies of transition metal atoms, ions, and monoxides using selected configuration interaction
and density-based basis-set corrections}

\author{Yuan Yao}
  \email{yy682@cornell.edu}
  \affiliation{Laboratory of Atomic and Solid State Physics, Cornell University, Ithaca, New York 14853, United States}
\author{Emmanuel Giner}
  \email{emmanuel.giner@lct.jussieu.fr}
  \affiliation{Laboratoire de Chimie Th\'eorique, Sorbonne Universit\'e and CNRS, F-75005 Paris, France}
\author{Tyler A. Anderson}
  \email{taa65@cornell.edu }
  \affiliation{Laboratory of Atomic and Solid State Physics, Cornell University, Ithaca, New York 14853, United States}
\author{Julien Toulouse}
  \email{toulouse@lct.jussieu.fr}
  \affiliation{Laboratoire de Chimie Th\'eorique, Sorbonne Universit\'e and CNRS, F-75005 Paris, France}
  \affiliation{Institut Universitaire de France, F-75005 Paris, France}
\author{C. J. Umrigar}%
  \email{cyrusumrigar@cornell.edu}
  \affiliation{Laboratory of Atomic and Solid State Physics, Cornell University, Ithaca, New York 14853, United States}

\begin{abstract}
The semistochastic heat-bath configuration interaction (SHCI) method is a selected configuration interaction plus perturbation theory method that has provided near-full configuration interaction (FCI) levels of accuracy for many systems with both single- and multi-reference character. However, obtaining accurate energies in the complete basis set limit is hindered by the slow convergence of the FCI energy with respect to basis size. Here we show that the recently developed basis-set correction method based on range-separated density-functional theory can be used to significantly speed up basis-set convergence in SHCI calculations. In particular, we study two such schemes that differ in the functional used, and apply them to transition metal atoms and monoxides to obtain total, ionization, and dissociation energies well converged to the complete-basis-set limit within chemical accuracy.

\end{abstract}

\maketitle

\section{Introduction}
\label{Intro}

Selected configuration interaction plus perturbation theory (SCI+PT)
methods~\cite{HurMalRan-JCP-73,GinSceCaf-CJC-13,Eva-JCP-14,SceAppGinCaf-JCoC-16,GarSceLooCaf-JCP-17,LooSceBloGarCafJac-JCP-18,HaiTubLevWhaHea-JCTC-19,LooLipPasSceJac-JCTC-20}
are capable of providing excellent approximations to full configuration interaction (FCI)
energies, i.e., the exact energies within a given one-electron basis, for systems and basis-sets that are considerably larger than those for which FCI is feasible.
In this paper, we employ a particularly efficient SCI method developed by some of the current authors,
namely the semistochastic heat-bath configuration interaction (SHCI) method \cite{HolTubUmr-JCTC-16,ShaHolJeaAlaUmr-JCTC-17,HolUmrSha-JCP-17,LiOttHolShaUmr-JCP-18}, which has been used for
a number of challenging problems, including the potential energy curve of the Cr$_2$ molecule~\cite{LiYaoHolOttShaUmr-PRR-20},
the dissociation energies of the 55 molecules comprising the G2 set~\cite{YaoGinLiTouUmr-JCP-20}
, and the ionization and dissociation energies of seven transition metal elements (their atoms, ions, and monoxides)~\cite{WilYao_etal_UmrWag-PRX-20}.

In particular, the SHCI results on transition metal systems have served as accurate benchmark energies for 19 other quantum chemistry methods for basis sets from DZ to 5Z, as well as the extrapolated complete-basis-set (CBS) limit \cite{WilYao_etal_UmrWag-PRX-20}.
However, unlike for the extrapolation to the FCI limit, whose extrapolation error is on the order of a couple of mHa for the largest monoxide molecules and much less for most other systems in the study, the largest extrapolation error to the CBS values can be on the order of 10 mHa
\footnote{The largest differences between the CBS345 and CBS45 values in Table V of the supplementary material of Ref.~\onlinecite{WilYao_etal_UmrWag-PRX-20} occur
for the heaviest system, CuO, and are 8 mHa.
However, the better extrapolation formula of Eq.~\ref{corr_extrap} in the current paper results in energies that are 6 mHa lower, and our best estimate
using the PBE-OT basis set correction (described later) followed by extrapolation results in a further 8 mHa lower energy.
The differences for the lighter transition metal systems are much smaller than this.}.
In addition, the computational cost goes up rapidly with basis size, so efficient techniques for accelerating
the convergence to the CBS limit become important in situations where the available computational resources limit the basis sizes one can use.

The slow convergence of SCI and other wave-function methods with respect to basis-set size
is due to the failure of wave functions
expanded in finite one-electron basis sets to reproduce the short-range correlation hole around the electron-electron cusp present in
exact wave functions.
F12 methods~\cite{Ten-TCA-12, HatKloKohTew-CR-12, KonBisVal-CR-12, GruHirOhnTen-JCP-17, MaWer-WIREs-18}
accelerate the basis-set convergence of wave function methods by restoring the electron-electron cusp with a correlation factor,
at the cost of introducing an auxiliary basis.
In particular, F12 methods have been used to accelerate the basis-set convergence of coupled-cluster calculations on transition metal systems~\cite{HilPet-JCTC-12,BroHilWerPet-JCP-13,KesMar-TCA-14,Tew-JCP-16}.
In this paper, we instead use two variants of a method based on range-separated density-functional theory (RS-DFT), recently
developed by some of the current authors~\cite{GinPraFerAssSavTou-JCP-18,LooPraSceTouGin-JPCL-19,GinSceTouLoo-JCP-19,GinSceLooTou-JCP-20,GinTraPraTou-JCP-21}.
The basic idea is that the electron-electron Coulomb interaction projected in a one-electron basis resembles
the long-range potential, $w^\text{lr}_\text{ee}(r_{12})= \text{erf}(\mu r_{12})/r_{12}$, used in RS-DFT.
The on-top value of the projected Coulomb interaction provides the local value of the range-separation parameter $\mu$ and functionals derived
in the context of RS-DFT can be used to provide a basis-set correction.

The outline of this paper is as follows.  In Section~\ref{SHCI_review} we briefly review the SHCI method.
In Section~\ref{basis_set_corr} we discuss the two basis-set correction schemes we employ in this paper.
Section~\ref{comput_details} contains the computational details.
In Section~\ref{results} we present the energies obtained with and without the two basis-set correction formulae.
The conclusions are in Section~\ref{conclusions}.

\section{Review of the SHCI method}
\label{SHCI_review}

In this section, we give a brief review of the SHCI method.

The variational stage of SHCI starts from an initial determinant such as the Hartree-Fock (HF) determinant
and generates the variational wave function, $\Psi_\text{V}$, through an iterative procedure.
At each iteration, $\Psi_\text{V}$ is written as a linear combination of the determinants
in the variational set $\V$
\begin{align}
\label{variational_wf}
\left|\Psi_\text{V} \right\rangle= \sum_{D_i \in \V} c_{i} \left|D_{i}\right\rangle,
\end{align}
and new determinants, ${D_a}$, from the perturbative set $\P$
(i.e., the determinants not in $\V$ but connected through the Hamiltonian to the determinants in $\V$) that satisfy the criterion
\beq
\exists\; D_i \in \V , \mathrm{\ such\ that\ } \left|H_{a i} c_{i}\right| \ge \epsilon_{1}
\label{HCI_criterion}
\eeq
are added to the set $\V$, where
$H_{ai}$ is the Hamiltonian matrix element between determinants $D_a$ and $D_i$, and
$\epsilon_1$ is a user-defined parameter that controls the accuracy of the variational
stage~\footnote{Since the absolute values of $c_i$ for the most important determinants tend to go down as more
determinants are
included in the wave function, a somewhat better selection of determinants is obtained by using a larger value of
$\epsilon_1$ in the initial iterations.}.
(When $\epsilon_1=0$, the method becomes equivalent to FCI.)
After adding the new determinants to the set $\V$, the Hamiltonian matrix is constructed and diagonalized using the diagonally
preconditioned Davidson method~\cite{Dav-CPC-89} to obtain an improved estimate of the lowest eigenvalue, $E_{V}$,
and eigenvector, $\Psi_\text{V}$.
This process is repeated until the change in the variational energy $E_\text{V}$ falls below a certain threshold.

Although the selection criterion in Eq.~(\ref{HCI_criterion}) is only an approximation to selection
criteria based on the largest contributions to the first-order perturbative correction to the wave function or the second-order perturbative correction to the energy,
the selected determinants are only slightly inferior to those selected by either of these perturbative criteria.
This is greatly outweighed by the improved selection speed.

In the perturbative stage of SHCI, a second-order Epstein-Nesbet perturbative energy correction $\Delta E^{(2)}$ is calculated
using the determinants connected to the final set $\V$ that satisfy Eq.~(\ref{HCI_criterion}) but with $\epsilon_1$ replaced by
$\epsilon_{2}$, which is typically orders of magnitude smaller than $\epsilon_{1}$:
	\begin{align}
		\Delta E^{(2)} =\; \sum_{D_a \in \P} \frac{\left(\sum_{D_i \in \V}^{(\epsilon_{2})} H_{ai} c_i\right)^2}{E_V - E_a},
		\label{eq:PTa}
	\end{align}
where $E_a = H_{aa}$ and $\sum^{(\epsilon_{2})}$ indicates that only terms in the sum for which $|H_{ai}c_i|\ge \epsilon_{2}$ are included.
The final SHCI energy is then $E_{\rm SHCI}=E_\text{V}+\Delta E^{(2)}$.

In both the variational and the perturbative stages, the fact that the number of distinct values of the double-excitation
matrix elements scales only as $N_{\rm orb}^4$, where $N_{\rm orb}$ is the number of orbitals, is used to avoid ever
looking at the very large number of unimportant determinants that do not contribute to the final energy.
Nevertheless, straightforward approaches for evaluating the perturbative correction entail either
storing all determinants of the perturbative set, $\P$, which creates a memory bottleneck for large systems,
or a significant increase in computer time.

SHCI circumvents this 
using a three-step semistochastic approach for the
second-order perturbation calculation, where the energy correction is split into deterministic, ``pseudo-stochastic'', and stochastic contributions \cite{LiOttHolShaUmr-JCP-18}.
The deterministic step chooses both variational and perturbative determinants deterministically,
the ``pseudo-stochastic'' step chooses variational determinants deterministically and
perturbative determinants stochastically, and the stochastic step chooses both variational and
perturbative determinants stochastically.
A threshold value $\epsilon_{2}^{\rm dtm}$ ($<\epsilon_{1}$), automatically determined to correspond to
a determinant set of manageable size depending on available computer memory,
is first used to obtain a deterministic energy correction.
The remaining correlation is calculated as the sum of a ``pseudo-stochastic" part obtained as
the difference in the second-order corrections evaluated with $\epsilon_{2}^{\rm dtm}$ and a smaller
threshold $\epsilon_{2}^{\rm psto}$, and a stochastic part obtained as the difference in corrections between
$\epsilon_{2}^{\rm psto}$ and the final threshold $\epsilon_{2}$ ($\epsilon_{2} < \epsilon_{2}^{\rm psto} < \epsilon_{2}^{\rm dtm} <\epsilon_{1}$, separated by one order of magnitude from one another).
Stochastic samples are taken until the statistical error falls below a specified threshold, which
is typically on the order of a few microHartree.

In a typical SHCI calculation, the variational energy and the corresponding perturbative correction are computed
for several values of $\epsilon_1$ (for the systems presented in this study, the lowest $\epsilon_{1}$ values are in the $2-4\times 10^{-5}$ range). To estimate the FCI energy, we perform a weighted quadratic fit of
$E_{\rm SHCI}$ to $-\Delta E^{(2)}$ to obtain $E_{\rm SHCI}$ at $-\Delta E^{(2)}=0$, using weights proportional
to $(\Delta E^{(2)})^{-2}$. In
order to reduce the extrapolation error, one can either go to larger variational wave functions by decreasing
$\epsilon_1$, incurring a large memory footprint, or optimize the
orbitals~\cite{YaoUmr-JCP-21} to minimize the variational energy for the same number of determinants.
To limit the computational cost of orbital optimization, the value of the threshold $\epsilon_1$ used during
optimization can be much larger than the value used to obtain the final energy.

\section{Basis-set correction}
\label{basis_set_corr}

One significant drawback of wave-function methods is the slow convergence of energies with respect to
the size of the one-electron basis set. This is because wave functions expanded in finite one-electron basis sets usually poorly describe short-range correlation near the electron-electron cusp.
To estimate the basis-set incompleteness error and speed up basis-set
convergence, we employ the density-based basis-set correction recently developed in
Refs.~\onlinecite{GinPraFerAssSavTou-JCP-18,LooPraSceTouGin-JPCL-19,GinSceTouLoo-JCP-19,GinSceLooTou-JCP-20,GinTraPraTou-JCP-21} by some of the current authors.

The starting idea of this basis-set correction scheme is that, in a given basis set $\mathcal{B}$, the exact ground-state energy $E_0$ of an $N$-electron system can be approximated by the energy $E_0^\mathcal{B}$ defined by the following minimization over $\mathcal{B}$-representable one-electron densities $n^{\mathcal{B}}({\bf r})$, i.e densities that can be obtained from a wave function $\Psi^\mathcal{B}$ belonging to the $N$-electron Hilbert space generated by the basis set $\mathcal{B}$,
\begin{equation}
\label{E0BminnB}
E_0^{\mathcal{B}}=\min_{n^{\mathcal{B}}}\left\{F[n^{\mathcal{B}}]+\int \text{d}{\bf r} \; v_{\rm ne}({\bf r}) n^{\mathcal{B}}({\bf r})\right\},
\end{equation}
where $v_{\rm ne}({\bf r})$ is the nuclei-electron potential and $F[n]=\min_{\Psi\rightarrow n}\langle \Psi | \hat{T}+\hat{W}_{\rm ee} | \Psi\rangle$ is the standard constrained-search Levy-Lieb universal density functional~\cite{Lev-PNAS-79,Lie-IJQC-83} with $\hat{T}$ and $\hat{W}_{\rm ee}$ being the kinetic and electron-electron Coulomb operators, respectively. Obviously, in the CBS limit, $E_0^{\mathcal{B}}$ tends to the exact ground-state energy, i.e. $\lim_{\mathcal{B}\to \text{CBS}} E_0^{\mathcal{B}} = E_0$.
Crucially, since the density usually has a fast convergence with the size of the basis set, the energy $E_0^{\mathcal{B}}$ also has a fast basis-set convergence.

As it stands, Eq.~(\ref{E0BminnB}) is of course not practical due to the lack of a sufficiently
accurate explicit approximation for the universal density functional $F[n]$.
However, as shown in Ref.~\onlinecite{GinPraFerAssSavTou-JCP-18}, by reintroducing a wave function,
$\Psi^{\mathcal{B}}$, the energy $E_0^\mathcal{B}$ can be expressed as
\begin{equation}
\label{E0BminPsiB}
E_0^{\mathcal{B}}=\min_{\Psi^{\mathcal{B}}}\left\{ \langle \Psi^{\mathcal{B}} | \hat{H} | \Psi^{\mathcal{B}} \rangle + \bar{E}^{\mathcal{B}}[n_{\Psi^{\mathcal{B}}}]\right\},
\end{equation}
where the minimization is over normalized $N$-electron wave functions $\Psi^{\mathcal{B}}$ expanded in the basis set $\mathcal{B}$, $\hat{H}$ is the total Hamiltonian, and $\bar{E}^{\mathcal{B}}[n]$ is a complementary basis-set correction density functional,
\begin{equation}
\label{eq:basis_corr}
 \bar{E}^{\mathcal{B}}[n] = \min_{\Psi \rightarrow n}  \brakett{\Psi}{\kinop + \weeop}{\Psi} -
 \min_{\psibasis \rightarrow n}  \brakett{\psibasis}{\kinop + \weeop}{\psibasis},
\end{equation}
which must be evaluated at the density of the wave function $\Psi^{\mathcal{B}}$. The basis-set correction density functional $\bar{E}^{\mathcal{B}}[n]$ compensates for the basis-set restriction on the wave function $\Psi^{\mathcal{B}}$ in Eq.~(\ref{E0BminPsiB}) and vanishes in the CBS limit. This scheme can be trivially generalized to a basis-set correction functional depending on spin-resolved densities $n_{\uparrow}({\bf r})$ and $n_{\downarrow}({\bf r})$ giving the same energy $E_0^{\mathcal{B}}$ as
\begin{equation}
\label{E0BminPsiBspin}
E_0^{\mathcal{B}}=\min_{\Psi^{\mathcal{B}}}\left\{ \langle \Psi^{\mathcal{B}} | \hat{H} | \Psi^{\mathcal{B}} \rangle + \bar{E}^{\mathcal{B}}[n_{\uparrow,\Psi^{\mathcal{B}}},n_{\downarrow,\Psi^{\mathcal{B}}}]\right\},
\end{equation}
where $\bar{E}^{\mathcal{B}}[n_{\uparrow,\Psi^{\mathcal{B}}},n_{\downarrow,\Psi^{\mathcal{B}}}]$ is a new basis-set correction functional evaluated at the spin-$\uparrow$ and spin-$\downarrow$ densities of the wave function $\Psi^{\mathcal{B}}$. Moreover, as shown in Ref.~\onlinecite{GinTraPraTou-JCP-21}, the basis-set correction scheme can also be extended to a functional depending on both the density $n({\bf r})$ and the on-top pair density $n_2({\bf r})$ giving again the same energy $E_0^{\mathcal{B}}$ as
\begin{equation}
\label{E0BminPsiBn2}
E_0^{\mathcal{B}}=\min_{\Psi^{\mathcal{B}}}\left\{ \langle \Psi^{\mathcal{B}} | \hat{H} | \Psi^{\mathcal{B}} \rangle + \bar{E}^{\mathcal{B}}[n_{\Psi^{\mathcal{B}}},n_{2,\Psi^{\mathcal{B}}}]\right\},
\end{equation}
where $\bar{E}^{\mathcal{B}}[n_{\Psi^{\mathcal{B}}},n_{2,\Psi^{\mathcal{B}}}]$ is yet another basis-set correction functional evaluated at the density and on-top pair density of the wave function $\Psi^{\mathcal{B}}$.


In practice, we use two approximate basis-set correction functionals. For the scheme of Eq.~(\ref{E0BminPsiBspin}), we use the PBE-UEG functional of the form~\cite{LooPraSceTouGin-JPCL-19}
\begin{equation}
\bar{E}^{\mathcal{B}}_\text{PBE-UEG}[n_\uparrow,n_\downarrow]=\int \text{d}{\bf r} \; f\left(n_\uparrow({\bf r}), n_\downarrow({\bf r}),\nabla n({\bf r}), \mu^{\mathcal{B}}({\bf r}) \right),
\end{equation}
where $f$ is a function of the spin-resolved densities $n_\uparrow({\bf r})$ and $n_\downarrow({\bf r})$, the density gradient $\nabla n({\bf r})$, and the local range-separation parameter $\mu^{\mathcal{B}}({\bf r})$. The function $f$ was constructed based on a short-range version~\cite{FerGinTou-JCP-19} of the Perdew-Burke-Ernzerhof (PBE)~\cite{PerBurErn-PRL-96} correlation functional. One key ingredient of this short-range functional is a parametrization (in terms of the spin-resolved densities) of the on-top pair density of the uniform electron gas (UEG)~\cite{GorPer-PRB-01,GorSav-PRA-06}, hence its name. Its full explicit form was given in Refs.~\onlinecite{LooPraSceTouGin-JPCL-19,GinSceTouLoo-JCP-19,GinSceLooTou-JCP-20,GinTraPraTou-JCP-21}.
For the scheme of Eq.~(\ref{E0BminPsiBn2}), we use the spin-unpolarized PBE-OT functional of the form~\cite{GinSceLooTou-JCP-20}
\begin{equation}
\bar{E}^{\mathcal{B}}_\text{PBE-OT}[n,n_2]=\int \text{d}{\bf r}\; g\left(n({\bf r}), \nabla n({\bf r}), n_2({\bf r}),\mu^{\mathcal{B}}({\bf r}) \right),
\end{equation}
where $g$ is a function of the density $n({\bf r})$, the density gradient $\nabla n({\bf r})$, the on-top pair density $n_2({\bf r})$, and the local range-separation parameter $\mu^{\mathcal{B}}({\bf r})$.
(To avoid confusion, we point out that the PBE-OT functional of the current paper was named the SU-PBE-OT functional in
Ref.~\onlinecite{GinSceLooTou-JCP-20}.)
The PBE-OT functional has essentially the same physical content as the PBE-UEG functional with the exception that it uses the on-top (OT) pair density $n_2({\bf r})$ of the wave function $\Psi^{\mathcal{B}}$ as an independent variable, instead of that of the UEG, which is more accurate for strongly correlated systems. Also, it uses the total density instead of the spin-resolved densities since the additional information provided by the spin-resolved densities is largely already contained in the on-top pair density $n_2({\bf r})$~\cite{GinSceLooTou-JCP-20}. Its full explicit form was given in Refs.~\onlinecite{GinSceTouLoo-JCP-19,GinSceLooTou-JCP-20,GinTraPraTou-JCP-21}.

The common key ingredient in these basis-set correction functionals is the local range-separation parameter $\mu^{\mathcal{B}}({\bf r})$ which provides a local measure of the incompleteness of the basis set, and is given by~\cite{GinPraFerAssSavTou-JCP-18,GinTraPraTou-JCP-21}
\beq
\label{localmu}
\mu^{\mathcal{B}}({\bf r}) = \frac{\sqrt{\pi}}{2} W^{\mathcal{B}}({\bf r}),
\eeq
where $W^{\mathcal{B}}({\bf r})$ is the on-top value of an effective two-electron interaction representing the Coulomb two-electron interaction in the basis set ${\mathcal{B}}$. In an orthonormal orbital basis $\{ \phi_p \}$ generated by the basis set $\mathcal{B}$, $W^{\mathcal{B}}({\bf r})$ is defined by
\beq
\label{Weff}
W^{\mathcal{B}}({\bf r}) =
\begin{cases}
\label{eq:WB}
 f^{\mathcal{B}}_{\Psi^{\mathcal{B}}_\text{loc}}({\bf r})/n_{2,\Psi^{\mathcal{B}}_\text{loc}}({\bf r}),    & \text{if $n_{2,\Psi^{\mathcal{B}}_\text{loc}}({\bf r}) \ne 0$,}
 \\
 +\infty, &
 \text{otherwise,}
\end{cases}
\eeq
with
\begin{gather}
\label{eq:fbasis_val}
f^{\mathcal{B}}_{\Psi^{\mathcal{B}}_\text{loc}}({\bf r}) = \sum_{pqrstu}^{\mathcal{B}}  \phi_p({\bf r}) \phi_q({\bf r}) V_{pq}^{rs}
\Gamma_{rs}^{tu} \phi_t({\bf r}) \phi_u({\bf r}),
\\
\label{eq:twordm_val}
n_{2,\Psi^{\mathcal{B}}_\text{loc}}({\bf r})
= \sum_{rstu}^{\mathcal{B}} \phi_r({\bf r}) \phi_s({\bf r}) \Gamma_{rs}^{tu} \phi_t({\bf r}) \phi_u({\bf r}),
\end{gather}
where the sums run over all the (occupied and virtual) orbitals generated by the basis set $\mathcal{B}$, $V_{pq}^{rs}=\langle pq | rs \rangle$ are the two-electron integrals, and $\Gamma_{rs}^{tu}$ is the (opposite-spin) two-body reduced density matrix associated with a localizing wave function $\Psi^{\mathcal{B}}_\text{loc}$.
The only purpose of the wave function $\Psi^{\mathcal{B}}_\text{loc}$ is to localize the otherwise nonlocal effective interaction obtained by projecting the Coulomb two-electron interaction in the basis set ${\mathcal{B}}$.
The local range-separation parameter $\mu^{\mathcal{B}}({\bf r})$ is very weakly dependent on this wave function $\Psi^{\mathcal{B}}_\text{loc}$ (the dependence on $\Psi^{\mathcal{B}}_\text{loc}$
of the numerator and denominator in Eq.~(\ref{Weff}) largely cancels)
and it should be thought of as essentially depending only on the basis set ${\mathcal{B}}$.
Importantly, in the CBS limit, the local range-separation parameter diverges, i.e. $\lim_{\mathcal{B} \to \text{CBS}} \mu^{\mathcal{B}}({\bf r})= +\infty$, independently of $\Psi^{\mathcal{B}}_\text{loc}$,
and in this case the form of the PBE-UEG and PBE-OT basis-set correction functionals ensures that they properly vanish, i.e. $\lim_{\mathcal{B} \to \text{CBS}} \bar{E}^{\mathcal{B}}_\text{PBE-UEG}[n_\uparrow,n_\downarrow] = 0$ and $\lim_{\mathcal{B} \to \text{CBS}} \bar{E}^{\mathcal{B}}_\text{PBE-OT}[n,n_2] = 0$.

Even though it is possible to perform the minimizations in Eqs.~(\ref{E0BminPsiBspin}) and~(\ref{E0BminPsiBn2}) using the PBE-UEG and PBE-OT basis-set correction functionals~\cite{GinTraPraTou-JCP-21}, in practice, for energy calculations, a non-self-consistent approximation can safely be used in which the basis-set correction functionals are added a posteriori to an approximation of the FCI energy calculated with the basis set $\mathcal{B}$. Here, as in our previous work~\cite{YaoGinLiTouUmr-JCP-20}, we use the total SHCI energy $E_\text{SHCI}^{\mathcal{B}}$ in the basis set $\mathcal{B}$. We then have two basis-set corrected SHCI energies.
First, we have the SHCI+PBE-UEG energy
\begin{equation}
\label{E0BSHCIPBEUG}
E_\text{SHCI+PBE-UEG}^{\mathcal{B}} = E_\text{SHCI}^{\mathcal{B}} + \bar{E}^{\mathcal{B}}_\text{PBE-UEG}[n_{\uparrow,\Phi^{\mathcal{B}}_\text{HF}},n_{\downarrow,\Phi^{\mathcal{B}}_{\text{HF}}}],
\end{equation}
where the PBE-UEG functional is evaluated at the spin-resolved densities obtained from the HF single-determinant wave function $\Phi^{\mathcal{B}}_{\text{HF}}$. We calculate the spin densities at the HF level only since the PBE-UEG functional is weakly dependent on the level at which the spin densities are calculated. Moreover, we choose also $\Psi^{\mathcal{B}}_\text{loc} = \Phi^{\mathcal{B}}_{\text{HF}}$ in Eq.~(\ref{eq:WB}) for calculating the local range-separation parameter $\mu^{\mathcal{B}}({\bf r})$ that enters into the PBE-UEG functional. Second, we have the SHCI+PBE-OT energy
\begin{equation}
\label{E0BSHCIPBEOT}
E_\text{SHCI+PBE-OT}^{\mathcal{B}} = E_\text{SHCI}^{\mathcal{B}} + \bar{E}^{\mathcal{B}}_\text{PBE-OT}[n_{\Psi^{\mathcal{B}}_\text{SHCI}},n_{2,\Psi^{\mathcal{B}}_\text{SHCI}}],
\end{equation}
where the PBE-OT functional is evaluated with the density and on-top pair density of the variational SHCI wave function ${\Psi^{\mathcal{B}}_\text{SHCI}}$ available at the end of the SHCI calculation. We use the SHCI wave function and not the HF single-determinant wave function because the PBE-OT functional is accurate only if it is evaluated at the on-top pair density obtained from an accurate multideterminant wave function. Moreover, we choose also $\Psi^{\mathcal{B}}_\text{loc} = \Psi^{\mathcal{B}}_{\text{SHCI}}$ for calculating the local range-separation parameter $\mu^{\mathcal{B}}({\bf r})$ that enters into the PBE-OT functional, even though the use of an accurate multideterminant wave function for $\Psi^{\mathcal{B}}_\text{loc}$ is not really important here. Comparing the two basis-set corrections, the PBE-UEG basis-set correction is simpler to compute since it uses the HF single-determinant wave function, but the PBE-OT basis-set correction which uses the on-top pair density from the SHCI wave function should be more accurate.

\section{Computational Setup}
\label{comput_details}

Our computations employ the eCEPP pseudopotentials of Trail and Needs~\cite{TraNee-JCP-17} and their
associated aug-cc-pV$n$Z-eCEPP basis sets with $n=2,3,4,5$.
These are abbreviated as DZ, TZ, QZ, and 5Z, or generically as $n$Z, in what follows.
The molecular geometries and the experimental  energies are the same as in the previous benchmark paper of
Ref. \onlinecite{WilYao_etal_UmrWag-PRX-20}.

We calculate the total energies in each of the four basis sets in three different ways:
\begin{enumerate}
 \item SHCI only. We first perform HF calculations with the software \textsc{PySCF}~\cite{sun_pyscf} and then
perform SHCI calculations with orbital optimization~\cite{YaoUmr-JCP-21} using the \textsc{Arrow} package \cite{Arrow_github}. These results have previously appeared in
 Ref. \onlinecite{WilYao_etal_UmrWag-PRX-20}.
 \item SHCI with PBE-UEG basis-set correction. The HF wave function is fed into \textsc{Quantum Package}~\cite{GarAppGasBenFerPaqPraAssReiTouBarRenDavMalVerCafLooGinSce-JCTC-19} to obtain the basis-set
 correction. The amplitude of the dominant determinant in the SHCI wave function ranges from 0.92 to 0.96 for
 the metal atoms and from 0.83 to 0.91 for the oxide molecules.
Therefore, we expect the HF spin-densities to be a reasonable approximation to the SHCI spin-densities.
The HF on-top pair density is not accurate, but the on-top pair density obtained from the UEG
that has the HF spin-densities is reasonably accurate.

 \item SHCI with PBE-OT basis-set correction. We perform SHCI variational calculations for several different values
 of the threshold $\epsilon_{1}$ and use the corresponding two-body reduced density matrices for the basis-set correction
 using \textsc{Quantum Package}~\cite{GarAppGasBenFerPaqPraAssReiTouBarRenDavMalVerCafLooGinSce-JCTC-19}.
 As the size of the variational wave function increases, the on-top pair density
 decreases, leading to decreasing magnitudes of the PBE-OT basis-set correction. We plot the basis-set correction versus $E_\text{V}$
 and use a quadratic function to extrapolate to the SHCI total energy limit, $E_\text{V}\rightarrow E_{\rm SHCI}$,
 which is then taken as the final PBE-OT basis-set correction for the given system and basis set.
\end{enumerate}

In two of the above three schemes, we also extrapolate the total energies to the CBS limit.
In earlier work by some of the present authors~\cite{WilYao_etal_UmrWag-PRX-20}, the
SHCI-only energies were extrapolated to the CBS limit
using separate extrapolations for the HF energy and the correlation
energy,~\cite{HelKloKocNog-JCP-97,HalHelJorKloKocOlsWil-CPL-98,HalHelJorKloOls-CPL-99}
\beq
\label{HF_extrap}
E_{\rm HF}^{\rm CBS} &=& E_{\rm HF}^{n\text{Z}} - a \exp{(-bn)}, \\
\label{corr_extrap_usual}
E_{\rm corr}^{\rm CBS} &=& E_{\rm corr}^{n\text{Z}} - c n^{-3}.
\eeq
In the current paper, the HF energy is still extrapolated in the same way using $n=3,4,5$, but the SHCI-only correlation energy is extrapolated using the expression
\beq
\label{corr_extrap}
E_{\rm corr}^{\rm CBS} &=& E_{\rm corr}^{n\text{Z}} - {c \over n^3+d},
\eeq
because we find it gives a far better fit for the pseudopotentials and associated basis sets used in this study.
Since Eq.~(\ref{corr_extrap_usual}) does not give a good fit, one would use only the larger basis sets $n=4,5$ to get a reasonable extrapolation, whereas with Eq.~(\ref{corr_extrap}) $n=2,3,4,5$ can
be used.
(Even using just $n=4,5$, Eq.~(\ref{corr_extrap_usual}) gives extrapolated energies that are
too high, as shown in Fig. 4 of the supplementary material.
For the heaviest system studied, CuO, the improved fit using Eq.~(\ref{corr_extrap}) and
$n=2,3,4,5$ makes the CBS total energy 6 mHa lower than the fit using Eq.~(\ref{corr_extrap_usual})
and $n=4,5$.
Detailed comparison of the various fits can be found in the supplementary material.)


In the SHCI+PBE-OT scheme, we use a single exponential function
\begin{align}
\label{pbe_extrap}
E_{\text{SHCI+PBE-OT}}^{\rm CBS} &=& E_{\text{SHCI+PBE-OT}}^{n\text{Z}} - a \exp{(-bn)}
\end{align}
with $n=3,4,5$, since we expect an exponential convergence of the SHCI+PBE-OT energy with respect to $n$ ~\cite{FraMusLupTou-JCP-15}.
(In the supplementary material we also show the SHCI+PBE-OT energies extrapolated to the CBS limit using $n=2,3,4$.)

Energies from the SHCI+PBE-UEG scheme are not extrapolated as the SHCI+PBE-UEG energies are nonmonotonic with basis size for many of the systems,
as will be shown in the next section.
Note however that our SHCI+PBE-UEG calculations on the G2 set~\cite{YaoGinLiTouUmr-JCP-20}
had a monotonic dependence on the basis size and so we were able to extrapolate those energies.

\section{Results}
\label{results}

\begin{table*}[tbp]
	\caption{Total, ionization (for the atoms), and dissociation energies (for the monoxide molecules) obtained from the three schemes in the CBS limit. The SHCI+PBE-OT values (emphasized in bold) are used as reference values in Section \ref{results} for comparison. Units: Ha.
	}
	\begin{ruledtabular}
		\begin{tabular}{l|rr>{\bf}r|rr>{\bf}r}
		\multicolumn{1}{c|}{} & \multicolumn{3}{c|}{Total energy} & \multicolumn{3}{c}{Ionization/Dissociation energy}\\
		\cline{2-7}
			System &	SHCI & SHCI+PBE-UEG & SHCI+PBE-OT &SHCI & SHCI+PBE-UEG &	SHCI+PBE-OT \\
			\hline
O      &  -15.8477  & -15.8490  & -15.8478  & 0.5000 & 0.5009 & 0.5004 \\
Sc     &  -46.4921  & -46.4966  & -46.4939  & 0.2409 & 0.2412 & 0.2411 \\
Ti     &  -58.0068  & -58.0117  & -58.0088  & 0.2508 & 0.2510 & 0.2510 \\
V      &  -71.2371  & -71.2412  & -71.2381  & 0.2473 & 0.2475 & 0.2475 \\
Cr     &  -86.8078  & -86.8111  & -86.8080  & 0.2488 & 0.2489 & 0.2489 \\
Mn     &  -104.1606 & -104.1648 & -104.1617 & 0.2726 & 0.2732 & 0.2730 \\
Fe     &  -123.7718 & -123.7784 & -123.7745 & 0.2901 & 0.2905 & 0.2903 \\
Cu     &  -197.6236 & -197.6388 & -197.6328 & 0.2832 & 0.2836 & 0.2839 \\[0.2cm]
O$^+$  &  -15.3477  & -15.3481  & -15.3475  & -- & -- & -- \\
Sc$^+$ &  -46.2511  & -46.2554  & -46.2528  & -- & -- & -- \\
Ti$^+$ &  -57.7560  & -57.7607  & -57.7578  & -- & -- & -- \\
V$^+$  &  -70.9898  & -70.9938  & -70.9907  & -- & -- & -- \\
Cr$^+$ &  -86.5591  & -86.5622  & -86.5591  & -- & -- & -- \\
Mn$^+$ &  -103.8879 & -103.8917 & -103.8887 & -- & -- & -- \\
Fe$^+$ &  -123.4817 & -123.4879 & -123.4843 & -- & -- & -- \\
Cu$^+$ &  -197.3404 & -197.3552 & -197.3489 & -- & -- & -- \\[0.2cm]
ScO    &  -62.5983  & -62.6040  & -62.6002  & 0.2585 & 0.2584 & 0.2585 \\
TiO    &  -74.1103  & -74.1168  & -74.1128  & 0.2557 & 0.2561 & 0.2562 \\
VO     &  -87.3302  & -87.3363  & -87.3318  & 0.2453 & 0.2461 & 0.2458 \\
CrO    &  -102.8299 & -102.8358 & -102.8316 & 0.1744 & 0.1758 & 0.1758 \\
MnO    &  -120.1529 & -120.1596 & -120.1550 & 0.1446 & 0.1458 & 0.1455 \\
FeO    &  -139.7769 & -139.7861 & -139.7814 & 0.1573 & 0.1587 & 0.1591 \\
CuO    &  -213.5821 & -213.5980 & -213.5903 & 0.1108 & 0.1102 & 0.1097 \\
		\end{tabular}
	\end{ruledtabular}
	\label{tab:reference_values}
\end{table*}

\begin{figure*}[htbp]
 \centerline{\includegraphics[width=1.15\textwidth]{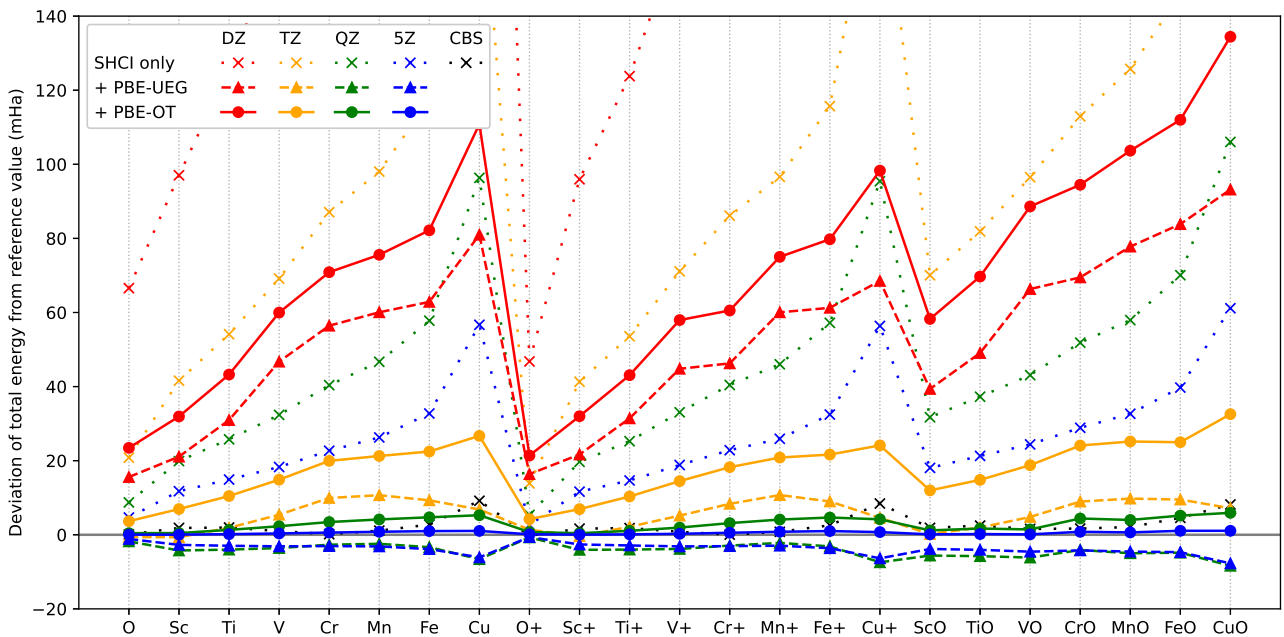}}
 \caption{Deviations of the total energies from reference (CBS SHCI+PBE-OT) values.
 Some of the SHCI-only values for the smaller basis sets are off the scale of the plot. }
 \label{fig:total_energy}
\end{figure*}

\begin{figure}[htbp]
	\centerline{\includegraphics[width=1.1\textwidth]{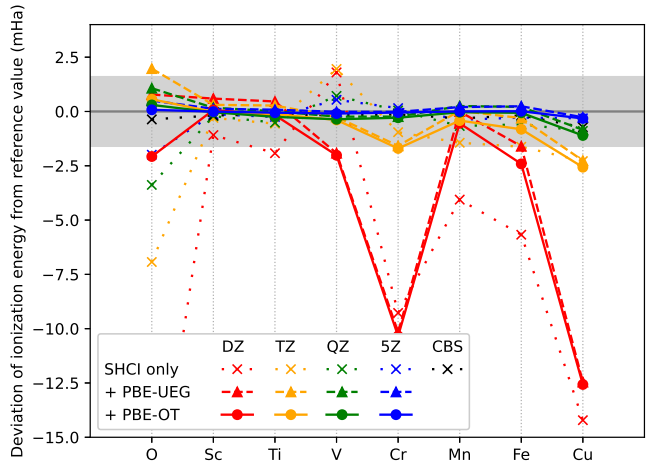}}
	\caption{Deviations of the ionization energies from reference (CBS SHCI+PBE-OT) values.
		The shaded area indicates chemical accuracy, i.e., 1 kcal/mol.}
	\label{fig:ionization_energy}
\end{figure}

\begin{figure}[htbp]
	\centerline{\includegraphics[width=1.15\textwidth]{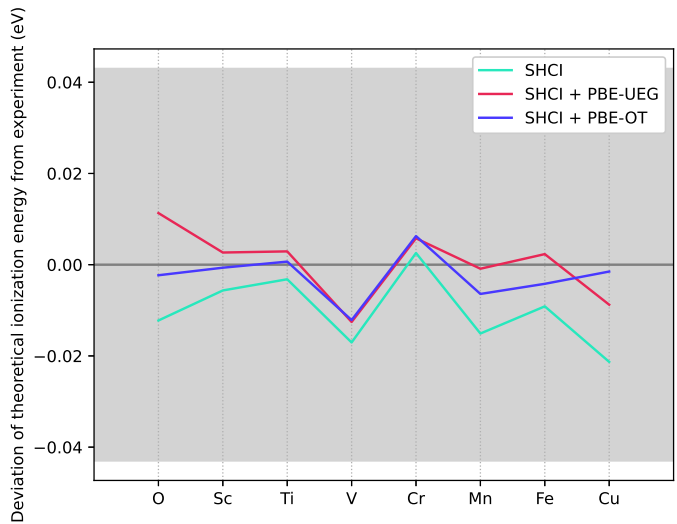}}
	\caption{Comparison of theoretical ionization energies with experiments. The theoretical values are obtained from the three schemes: SHCI only (CBS), SHCI+PBE-UEG (5Z), and SHCI+PBE-OT (CBS). The shaded area indicates chemical accuracy, i.e., 1 kcal/mol.}
	\label{fig:experiment_ionization}
\end{figure}

\begin{figure*}[htbp]
 \centerline{\includegraphics[width=1.15\textwidth]{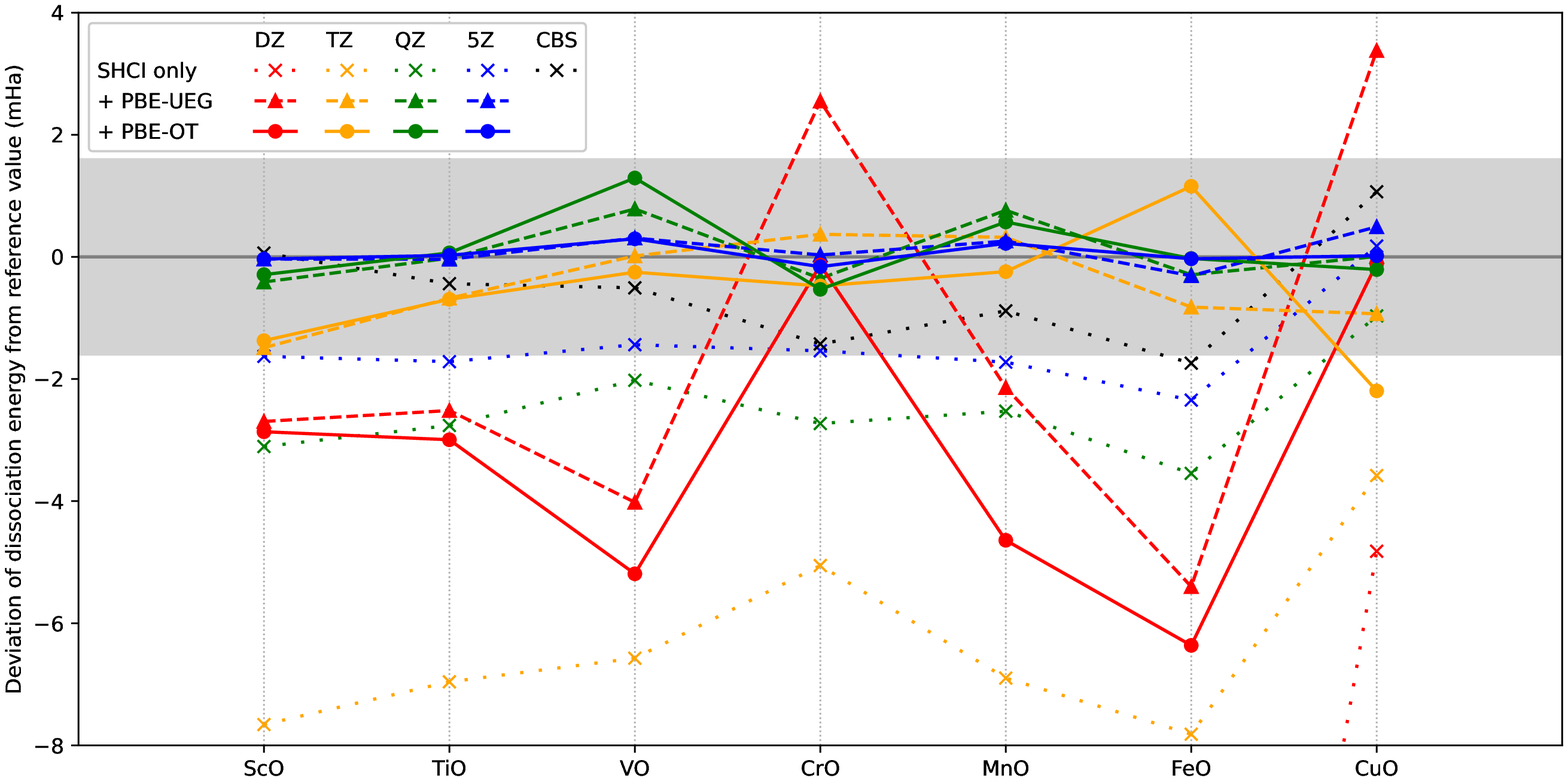}}
 \caption{Deviations of the dissociation energies from reference (CBS SHCI+PBE-OT) values.
 The shaded area indicates chemical accuracy, i.e., 1 kcal/mol.
 Some of the SHCI-only values for the smaller basis sets are off the scale of the plot.}
 \label{fig:dissociation_energy}
\end{figure*}

In this section, we present energies for the seven monoxide molecules, the eight constituent atoms, and the corresponding ions using
the three schemes described in the previous section. The results are shown for each of the four basis sets  DZ, TZ, QZ,
and 5Z, as well as for the extrapolated CBS limit.
The total energies, first ionization energies (for the eight atoms) and dissociation energies (for the seven monoxide molecules) are reported in Table~\ref{tab:reference_values}.

In what follows, we use as reference values the CBS results obtained
from the SHCI+PBE-OT scheme.
We think it likely that this scheme gives the most accurate energies since the SHCI+PBE-OT scheme
employs a more accurate on-top pair density than the SHCI+PBE-UEG scheme.
In the supplementary material, the quantum Monte Carlo (QMC) energies
for the oxygen atom also corroborate
the selection of SHCI+PBE-OT as the reference.
However, we also acknowledge that this choice of reference values is not certain.
In our earlier work on the G2 set of 55 molecules, we employed only the SHCI-only and the
SHCI+PBE-UEG schemes since those molecules are sufficiently weakly correlated.
There we found that although for each of the finite basis sets the SHCI+PBE-UEG energies agreed
considerably better with experiment than the SHCI-only energies, the CBS energies from
SHCI-only gave slightly better agreement than those from SHCI+PBE-UEG.

\subsection{Total Energies}

Figure~\ref{fig:total_energy} shows deviations of total energies from the reference values. Compared to the SHCI-only
results, both basis-set correction schemes significantly  reduce the basis  incompleteness error.
For each system in a particular basis, the correction from PBE-UEG is consistently larger in magnitude than from
PBE-OT. This makes the PBE-UEG-corrected values closer to the CBS limit for smaller basis sets, but for larger
basis sets PBE-UEG tends to overcorrect. In fact, for many of the systems shown here the PBE-UEG-corrected energies
for 5Z are higher than for QZ, reversing the trend from DZ to QZ. Similar nonmonotonicity of the corrected values
with basis size has been observed in previous studies as well~\cite{GinTraPraTou-JCP-21}.

We have separately verified the total energy in the CBS limit by performing QMC calculations on the oxygen atom (see the supplementary material).
The QMC total energy is in good agreement with the CBS SHCI-only and SHCI+PBE-OT energies and differ
considerably from the SHCI+PBE-UEG energy.
The energy difference between the former two schemes may be viewed as an estimate of the uncertainty in
the CBS energies.  Table \ref{tab:total_energy} reports the mean absolute
deviations (MADs) from the reference values for the different basis sets and methods.

\begin{table}[htbp]
 \caption{Mean absolute deviations of total energies from reference (CBS SHCI+PBE-OT) values for different basis sets. Units:
 mHa.
 }
\begin{ruledtabular}
\begin{tabular}{lrrrrr}
 {} &	    DZ &	TZ &	    QZ &	5Z &	CBS \\
\hline
SHCI             & 217.67 & 95.04 & 45.58 & 26.06 &    2.47 \\
SHCI + PBE-UEG   &  52.35 &  5.63 &  4.21 &  3.70 &     -- \\
SHCI + PBE-OT    &  70.75 & 17.37 &  2.87 &  0.51 &    0.00 \\
\end{tabular}
\end{ruledtabular}
\label{tab:total_energy}
\end{table}

\subsection{Ionization Energies}
Figure \ref{fig:ionization_energy} shows the convergence of the first ionization energies of the eight atoms with respect to basis size.
For the metal atoms, although the ionization energies obtained with the DZ basis sets for Cr and Cu show large errors,
all three schemes are almost converged to within chemical accuracy at the level of TZ,
and converged to far better than chemical accuracy at the level of 5Z.
The MADs are reported in Table \ref{tab:ionization_energy}.
Most of the MADs are more than an order of magnitude smaller than those for the total energies
because of a large cancellation of error between the atomic and ionic energies.

Accurate experimental ionization energies are also available for these systems.
In Fig. \ref{fig:experiment_ionization}, we compare the theoretical ionization energies obtained from the three schemes (i.e. extrapolated energies from the SHCI-only and SHCI+PBE-OT schemes, and the 5Z energies from the SHCI+PBE-UEG scheme) to experiment.
Much better agreement than chemical accuracy is obtained for all the atoms and all three schemes.
Remarkably the largest deviation of the energies obtained from either of the two correction schemes and experiment
is only 0.01 eV.
The MADs from experiment are 10.8, 5.9, and 4.3 meV for SHCI-only, SHCI+PBE-UEG and SHCI+PBE-OT, respectively.
The high level of agreement of the SHCI+PBE-OT energies with experiment provides further support for using SHCI+PBE-OT energies as reference values,
and speaks to the accuracy of the experiments as well.

\begin{table}[htbp]
	\caption{Mean absolute deviations of ionization energies from reference (CBS SHCI+PBE-OT) values for different basis sets. Units:
		mHa.
	}
\begin{ruledtabular}
\begin{tabular}{lrrrrr}
	{} &   DZ &   TZ &   QZ &   5Z &  CBS \\
	\hline
SHCI             & 7.23 & 2.01 & 0.88 & 0.50 &    0.28 \\
SHCI + PBE-UEG   & 3.50 & 0.87 & 0.38 & 0.18 &     --  \\
SHCI + PBE-OT    & 3.77 & 0.82 & 0.30 & 0.08 &    0.00 \\
\end{tabular}
\end{ruledtabular}
\label{tab:ionization_energy}
\end{table}

\subsection{Dissociation Energies}
Fig.~\ref{fig:dissociation_energy} shows the dissociation energies in the different basis sets for the three  schemes. As
expected, the basis incompleteness errors in energy differences are  around one order of magnitude smaller than
in the individual total energies themselves. For both basis-set correction schemes, at the level of the TZ basis set almost all systems
are converged to within chemical accuracy (with the single exception of CuO in the PBE-OT scheme).
Table \ref{tab:dissociation_energy} reports the MADs for the different basis sets and
methods.

\begin{table}[htbp]
 \caption{Mean absolute deviations of dissociation energies from reference (CBS SHCI+PBE-OT) values for different basis sets. Units:
 mHa.
 }
 \begin{ruledtabular}
 	\begin{tabular}{lrrrrr}
 		{} &	    DZ &	TZ &	    QZ &	5Z &	CBS \\
 		\hline
SHCI             & 15.52 & 6.36 & 2.52 & 1.51 &    0.88 \\
SHCI+PBE-UEG   &  3.24 & 0.66 & 0.38 & 0.21 &      -- \\
SHCI+PBE-OT &  3.18 & 0.91 & 0.43 & 0.11 &    0.00 \\
 	\end{tabular}
 \end{ruledtabular}
 \label{tab:dissociation_energy}
\end{table}

\begin{figure}[h!]
	\centerline{\includegraphics[width=0.99\textwidth]{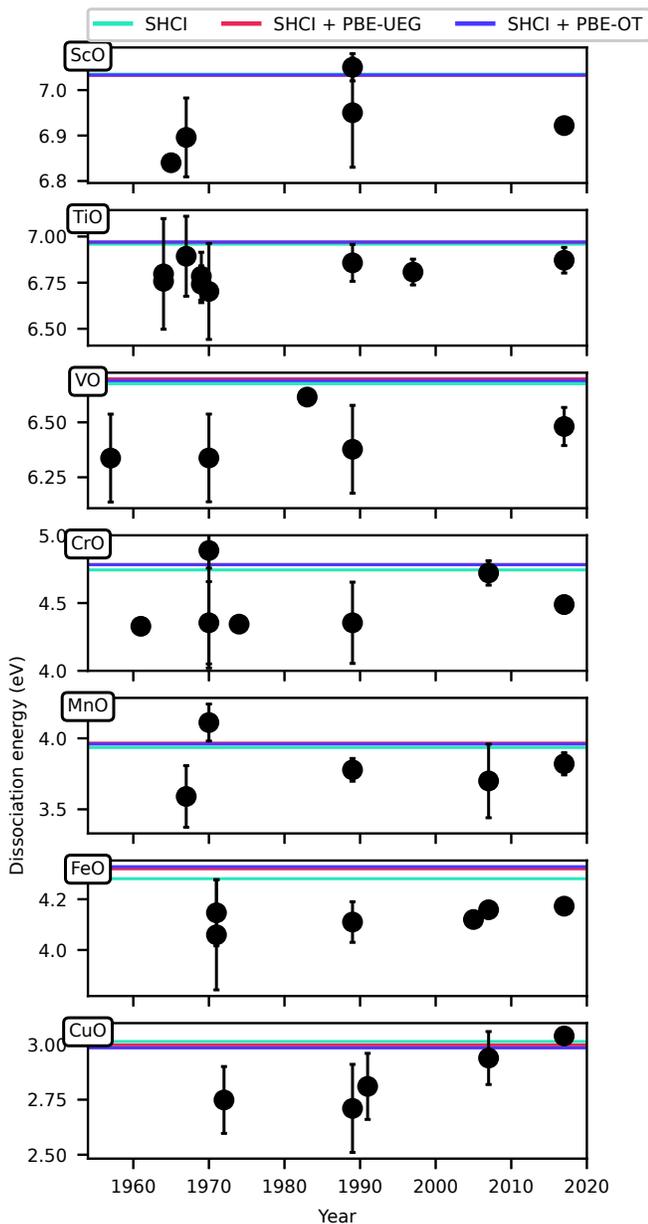}}
        \vspace{-9mm}
	\caption{Comparison of theoretical dissociation energies with experiments. The theoretical values obtained from the three schemes---SHCI only (CBS), SHCI+PBE-UEG (5Z), and SHCI+PBE-OT (CBS)---are converged to much higher precision than the experimental uncertainties.}
	\label{fig:experiment}
\end{figure}

In Fig. \ref{fig:experiment}, we take the final dissociation energies obtained from the three schemes
 and compare to experimental values reported over the years (see Ref. \onlinecite{WilYao_etal_UmrWag-PRX-20} and references therein).
These experimental values have been corrected for zero-point energy contributions but not for relativistic effects since
the pseudopotentials we use incorporate scalar-relativistic effects.
Unlike the ionization energies, the  experimental dissociation energies have large uncertainties and
differ significantly among themselves. For all systems studied, the theoretical values from our three schemes agree with each other to much higher precision than the
differences between the various experimental values.

\section{Conclusions}
\label{conclusions}

In prior work~\cite{WilYao_etal_UmrWag-PRX-20} we have shown that the SHCI method can be used to calculate near exact
energies for transition-metal atoms, ions, and oxides in basis sets up to 5Z.  There, the SHCI energies were used
as the reference for testing the accuracy of 19 other electronic-structure methods for each of the basis sets.
The CBS energies were also estimated by extrapolation using the formulae in Eqs.~(\ref{HF_extrap}) and
(\ref{corr_extrap_usual}).  In the current paper we have shown that a more accurate extrapolation is possible
using Eqs.~(\ref{HF_extrap}) and~(\ref{corr_extrap}).
More importantly, we have shown that density-based basis-set corrections can be used to estimate the basis incompleteness error of a finite basis set and speed up convergence to the CBS limit.
Two different schemes have been applied, namely, PBE-UEG and PBE-OT, differing in the functional used in the calculation.
Both methods produce ionization and dissociation energies converged to within chemical accuracy of the CBS limit with only the TZ basis set, which avoids the high cost of going to larger basis sets using a wave-function method such as SHCI.
We have also compared the ionization and dissociation energies from the two correction schemes as well as the uncorrected extrapolations to experimental values reported over the years.
Remarkably, for ionization energies, the largest deviation of the experimental energies from the energies obtained by either of the two correction schemes
is only 0.01 eV.
Our computed dissociation energies are converged to much higher precision than the experimental uncertainties.

\section*{Supplementary Material}
See supplementary material for detailed comparisons of basis-set extrapolations for SHCI total energies and SHCI+PBE-OT energies, as well as QMC results on the oxygen atom.

\vspace{5mm}
\begin{acknowledgements}
Y.Y. acknowledges fellowship support from the Molecular Sciences Software Institute funded by U.S. National Science
Foundation (Grant No. ACI-1547580).
This work was supported in part by the AFOSR (Grant No. FA9550-18-1-0095).
Some of the computations were performed at the Bridges cluster at the Pittsburgh Supercomputing Center supported
by the NSF (Grant No. ACI-1445606).
\end{acknowledgements}

\section*{Data Availability}
The data that support the findings of this study are available within the article and its supplementary material.

\FloatBarrier

\bibliographystyle{apsrev4-2}
\bibliography{all}

\end{document}